Bell's inequalities I: an explanation for their experimental violation

Louis Sica

Code 5630
Naval Research Laboratory
Washington, D. C. 20375
U.S.A.

Ph: 202-767-9466
Fax:202-767-9203
sica@ccf.nrl.navy.mil

Abstract

Derivations of two Bell's inequalities are given in a form appropriate to the interpretation of experimental data for explicit determination of all the correlations. They are arithmetic identities independent of statistical reasoning and thus cannot be violated by data that meets the conditions for their validity. Two experimentally performable procedures are described to meet these conditions. Once such data are acquired, it follows that the measured correlations cannot all equal a negative cosine of angular differences. The relation between this finding and the predictions of quantum mechanics is discussed in a companion paper.

The present work [1] (a preliminary version of some of the basic ideas contained in this paper was presented in Ref. [1]) divides consideration of Bell's inequalities and their widely believed violation by quantum mechanics into two separate domains: that of literally performable experimental tests, and that of gedanken experiments that depend on additional theoretical assumptions. Bell's original description of his inequality [2-4] was of the latter type: a thought experiment, involving theoretical constructs, that cannot actually be carried out. The issues raised by these contrasting viewpoints lead to important insights into the subject of Bell's inequalities violation. The subject of purely experimental tests is treated in this paper, Part I. The discussion of Bell's gedanken experiment is given in a companion paper Part II [5].

The vehicle for resolving some of the paradoxes of Bell's inequality violation is the distinction between what will here be called Bell's identities, and Bell's inequalities. Several authors have derived Bell's identities in passing on the way to derivation of Bell's inequalities, but without deducing their full implications for the understanding of Bell's inequalities paradoxes. Among these are Eberhard [6] (data matching is discussed in this work but all the logical implications in the context of Bell's identities are not realized), Peres [7], and Redhead [8] (data matching is discussed at length, but without realizing its full logical consequences).

Since derivations of Bell's identities are short, and critical to the understanding of the arguments that follow, they are given here, both for Bell's original three-correlation case, and for the four-correlation case [9]. It is useful to state their salient characteristics in advance: Bell's identities are properties of lists of numbers, restricted to $\pm 1$, that must be satisfied regardless of any other deterministic or statistical attributes. Bell's inequalities are limits, as list length goes to infinity, of the correlations among the numbers of the lists. If computed theoretically, the formulas for these correlations can violate the inequalities, since the inequalities are not identically satisfied by such formulas. However, if the inequalities are violated, then the derivation of the correlations must contain an error, since the streams of ± 1's that they represent must identically satisfy the prior identities. These ideas can be used to unravel some of the paradoxes of Bell's inequality violation.

The derivation of Bell's identity for three correlations is obtained first. Let us assume that we have three lists of numbers, each of length $N$, with each number restricted to the values ±1. The lists are denoted $a$, $b$, and $b'$ and their respective members by $a_i$, $b_i$, and $b_i'$, i = 1...N. The following equation is computed from corresponding members of the three lists for each value of $i$:

$$a_i b_i - a_i b_i' = a_i (b_i - b_i') = a_i b_i (1 - b_i b_i') \tag{1a}$$

By summing this equation over the list, dividing by $N$, and taking absolute values, one obtains:

$$\left| \sum_{i=1}^{N} a_i b_i / N - \sum_{i=1}^{N} a_i b_i' / N \right| \le \sum_{i=1}^{N} |a_i b_i| |1 - b_i b_i'| / N = \sum_{i=1}^{N} (1 - b_i b_i') / N \tag{1b}$$

or finally:

$$\left| \sum_{i=1}^{N} a_i b_i / N - \sum_{i=1}^{N} a_i b_i' / N \right| \le 1 - \sum_{i=1}^{N} b_i b_i' / N \tag{2}$$

Eq. 2 will be referred to as Bell's three-correlation identity to make absolutely clear that if three lists of numbers with values restricted to ±1 exist, they immediately and identically satisfy this relation independently of the size of $N$.

Eq. (2) is closely related to Bell's inequality. Since the truth of (2) is independent of $N$, then $N$ may be allowed to approach infinity, and if limits exist for the sums in (2), the limits must satisfy (2) also. Thus, in the limit, (2) becomes Bell's inequality

$$|<ab> - <ab>| \quad 1 - <bb> \tag{3}$$

where the angular brackets denote the limits of the corresponding sums in (2). It is perhaps startling to the reader to find that Bell's inequality has been obtained here with no mention of locality or nonlocality, properties of probability functions, factorization assumptions, etc. All these assumptions are peripheral to the central fact: identity (2) and inequality (3) follow from nothing but the arithmetic of ± 1's and the assumption of limits.

The concern of this paper will be primarily Bell's identity (2), and not inequality (3). This is because all experiments deal with finite data streams, even if those data streams are very large. Important consequences immediately follow from this fact. First, once three data streams have been obtained, (2) is immediately satisfied. Only two questions remain: how to obtain three data streams, and what correlations may result among them.

In applying (2) to observations, three variables must be logically identified from the physical situation. In the measurement of spin components of two spin-1/2 particles flying apart in a singlet state (Fig 1), the apparatus is assumed to be run in a delayed choice mode with angular settings of Stern-Gerlach magnets made on the fly. The treatment of polarization measurements for the pairs of photons which have been the subject of most experiments in this field is very similar, see for example, Ref. [10]. The *a* and *b* measuring devices are separated by a distance large compared to that which light can traverse in the time between measurements. The *a* measurement is completed before the *b* measurement is begun. This condition is relevant to reasoning employed in the gedanken experiments described in Part II. It is not, however, necessary for the employment of the three correlation inequality: the measurements could be carried out simultaneously. For a given orientation of the magnetic field *a*, the field at *b* may also be set in a different orientation *b'* resulting in a different stream of data. The data are random streams of ± 1's for this physical situation.

In such an experiment, two particles in a singlet state yield one number each so that only one correlation can be computed for *N* pairs of particles at fixed angular settings. This leads to important and unavoidable difficulties in the application of Bell's inequalities to experiments. The derivation of Bell's identity and inequality given above, as well as all other derivations known to the author, contain the factoring step contained in Eq. (1a). This implies that the same value for *a* must multiply both *b* and *b'* for each realization of the three variables. This requirement cannot be ignored if logical consistency with the derivation is to be maintained. Only two methods for realizing such data sets from Bell-pairs of particles are known to the author: data matching from successive experiments, or adding another Stern-Gerlach apparatus at each end of the train to realize a complete set of outputs for each pair of particles. Neither conforms to the gedanken experiment proposed by Bell discussed in Part II, but each provides some illumination of the issues.

In the data matching scheme, another set of particles must be measured at settings *a* and *b'* to obtain a data stream at *b'*. However, these measurements will result in different values for variable *a* from the first set, although the average number of + 1's and − 1's will be equal in both sets. To remedy this difficulty, the data of one of the sets must be reordered so that the sequence of values for *a* is the same in both. Since the probability for either *a* = + 1 or *a* = - 1 equals 1/2, then for increasingly long runs it should be possible to get a complete match of *a*-values while dropping a smaller and smaller percentage of data pairs (0% in the limit of infinite $N$ according to the law of large numbers). This procedure would have no effect on <*ab*> or <*ab'*> in the large $N$ limit, since a mean is independent of the order of addition of the data. A reordering of *b'* values is envisioned in which their randomness is maintained except for their conditional dependence on the *a*s. The result is that <*bb'*> may be computed from the data which is *now* consistent with the derivation of Bell's inequality, since the four data streams from two runs have effectively been reduced to three streams by the reordering of data from one run. Of course (2) is identically and trivially satisfied and its satisfaction does not depend on the amount of data taken. It will be satisfied by even small amounts of data for which the experimental correlations may be far from their limits. In this case, the fluctuation of the correlations will occur together since (2) is an arithmetic constraint independent of statistics.

The only question remaining is what the correlations are. For large amounts of data, the experimental correlations may closely approximate ensemble averages <*ab*>

and <ab'>, as well as <bb'> if the limits exist. It is of course widely known that <ab> and <ab'> are given by quantum mechanics as the negative cosine of angular differences. This would be expected to be the case for the data processing suggested here, but it cannot also be true for <bb'> at all angles, since as pointed out by Bell [2-4], (3) is then violated for some choices of angles. Consequently, it may be predicted that if a three correlation experiment were to be performed, the measured value of <bb'> would be different from that assumed by Bell. It should additionally be observed that even though *b* and *b'* do not commute quantum mechanically, <bb'> cannot always equal zero since for some angles the sum of the terms within the absolute value sign of (3) may be made larger than 1. Consequently, the existence of the data streams in this case requires <bb'> ≠ 0. From the point of view of random processes, the probability for each of the two possible values of *b* and *b'* is conditional on that of *a*, so that the correlation <bb'> is defined for this experiment even though *b* and *b'* do not commute. The experiment is analogous to Bell's gedanken experiment (to be discussed in Part II) in which a measurement procedure is thought of as being repeated with the same sequence of hidden variable values but different apparatus settings, and from which the sets of output values may then be correlated.

A similar situation holds for the four correlation version of Bell's inequality first derived by Clauser et al. [9]. The Bell's identity for four correlations follows immediately from the assumption that there exist four lists of numbers, restricted to +1 or - 1, of length *N*. The following two equalities are computed from the *i*th members of the lists *a, a', b, b'*:

$$a_i b_i + a_i b_i = a_i(b_i + b_i)  \qquad (4a)$$

and

$$a_i b_i - a_i b_i = a_i (b_i - b_i) \ . \qquad (4b)$$

Expressions (4a) and (4b) may be summed over the *N* members of the lists and absolute values taken resulting in two inequalities. These are added together and the absolute value brought inside the sum on the right-hand-side to obtain:

$$\left| \frac{1}{N} \sum_{i=1}^{N} a_i b_i + \frac{1}{N} \sum_{i=1}^{N} a_i b_i \right| + \left| \frac{1}{N} \sum_{i=1}^{N} a_i b_i - \frac{1}{N} \sum_{i=1}^{N} a_i b_i \right|$$

$$\leq \frac{1}{N} \sum_{i=1}^{N} |a_i| |b_i + b_i| + \frac{1}{N} \sum_{i=1}^{N} |a_i| |b_i - b_i| \ . \qquad (5)$$

The right-hand-side of (5) may be shown to be equal to 2. Therefore, Bell's identity for the four correlation case is

$$\left| \frac{1}{N} \sum_{i=1}^{N} a_i b_i + \frac{1}{N} \sum_{i=1}^{N} a_i b_i \right| + \left| \frac{1}{N} \sum_{i=1}^{N} a_i b_i - \frac{1}{N} \sum_{i=1}^{N} a_i b_i \right| \leq 2 \qquad (6)$$

independently of the magnitude of *N*. As before, Bell's identity is closely related to Bell's inequality. If *N* is allowed to approach infinity, (6) may be replaced by Bell's inequality

$$|\langle ab \rangle + \langle ab \rangle| + |\langle a b \rangle - \langle a b \rangle| \leq 2 \qquad (7)$$

if the limits of the sums exist. Identity (6) will be of concern here rather than (7) due to the finite character of experimental data sets. Once four data streams are in hand that satisfy the meager requirements of the derivation, there can be no violation of (6).

These four data streams can be obtained in a manner analogous to that prescribed above for three data streams. First, measure *a* and *b*, then *a* and *b'*. Now reorder the latter set until the *a*-data streams from the two runs match, line by line. From the law of large numbers, as N -> , the fraction of numbers equal to + 1 or - 1 approaches 1/2 for each list. Finally, measure *a'* and *b*, and reorder the list until the *b*-data stream matches that of the first *a* - *b* run. Data is now in hand to approximate not only <*ab*>, <*ab'*>, and <*a'b*>, but in addition, <*a'b'*>. If a separate run is carried out to measure <*a'b'*>, the data will be numerically over determined since the four data streams required by the derivation are already in hand from prior runs. If data from separate runs are used to compute correlations without reordering or attention paid to extra data streams, strict logical connection to the derivation of Bell's identity and inequalities is severed. In the classic experiment of Aspect et al. [11], separate runs are used for each correlation. Such a procedure involves an additional assumption about the nature of the stochastic process described by the correlations that will be examined in Part II. Note that violation of the conditions for the derivation of Bell's identity (6) and inequality (7) does not imply that the corresponding correlations violate Bell's inequality or that they do not. However, violation of the inequality by the correlations does imply that they cannot represent any data streams that could possibly exist or be imagined.

While it would appear that the first three correlations are expected to be equal (in the limit of large *N*) to the negative cosine of angular differences of detector settings, the corresponding value for <*a'b'*> will not be so equal, as it is known that for this combination of correlation functions, Bell's inequality is violated for certain sets of angles. Therefore, no data streams having those correlations can exist. However, it may be observed that since the correlation $<a'b'>$ is clearly statistically conditional on the others, it is not surprising that it must be functionally different from them.

The schemes for applying Bell's identities to experimental data envisioned here are different from that implied by Bell's derivation. His derivation and associated gedanken experiment, to be considered in Part II, envisioned computing one correlation from one stream of particle pairs and constructing the other correlations from theoretical reasoning. The data stream matching scheme seems to be the only way to obtain data for all the correlations using the apparatus he considered. However, by augmenting the conventional arrangement as shown in Fig. 2, four spin values may be obtained from each particle pair by retrodiction. (That back-calculation/retrodiction can furnish values for non-commuting variables was known to Heisenberg, see Ref. [12].) By placing an additional Stern-Gerlach apparatus on each side beyond the conventional ones, two spin measurements per particle may be inferred from the final output spot. Such apparatus would yield data streams to form either three or four correlations that would satisfy Bell's identities, but would yield correlations different from those obtained with the data stream matching method. These correlations have been computed and of course are different from the single cosine functions usually assumed for Bell's inequalities correlations.

The present paper has shown that the form of Bell's inequalities appropriate for comparison with experiments is an identity based on minimal assumptions. If these minimal assumptions are not met by the data, the conditions for validity of the identity will be violated, and the inequality may (or may not) be violated. This may happen if it is not noticed that the data for two correlation estimates uniquely determines the third in the three correlation case, and that data for three correlations determine the fourth in the four correlation case. That this has not been generally recognized may stem from the belief, based on early derivations, that Bell's inequality is a fact about statistics. But in fact, as has been shown above, it is a constraint of arithmetic, quite independently of statistics.

In the most common interpretation of Bell's inequalities experiments, only one correlation is literally measured while the others are inferred from quantum mechanics together with other assumptions. The violation of the inequality implies that these additional assumptions are somehow flawed, as no data streams consistent with arithmetic could exist giving rise to the correlation actually observed together with those theoretically inferred. Of course, this leaves open the question of what values the inferred correlations should take, assuming that the inferred data exists. The issue of whether this finding violates quantum mechanics in some sense will be discussed in Part II, the companion to this paper.

Acknowledgements

I gratefully acknowledge Michael Steiner for many stimulating conversations on the subject of this article and for pointing out a number of useful references. I am also indebted to Yanhua Shih and Carlton M. Caves for critical comments on an early version of the manuscript, and to Kent S. Wood for useful suggestions regarding presentation of the material. Finally, I would like to thank Herschel S. Pilloff for continued help and encouragement during the course of this study.

Figures

1. Schematic Stern-Gerlach apparatus. Arrows a, b, indicate magnetic field directions encountered by pairs of particles emitted in opposite directions by the source. Distances from the source to the measuring devices are shown unequal to allow the a-measurement to be performed first. At each encounter with a magnetic field, the particle is deflected in one of two directions depending on whether its spin is +1 or -1 (half). The direction b' indicates an alternative direction of b for which data would result if that direction were selected.

2. Schematic Multiple Stern-Gerlach apparatus. Arrows a, a', b, b' indicate the magnetic field directions encountered by pairs of particles emitted in opposite directions by the source. At each encounter with a magnetic field, the particle is deflected in one of two directions depending on whether its spin is +1 or -1 (half). Each sequence of 1's corresponds to a unique spot position so that knowledge of two spin measurements is yielded retrospectively for each particle.

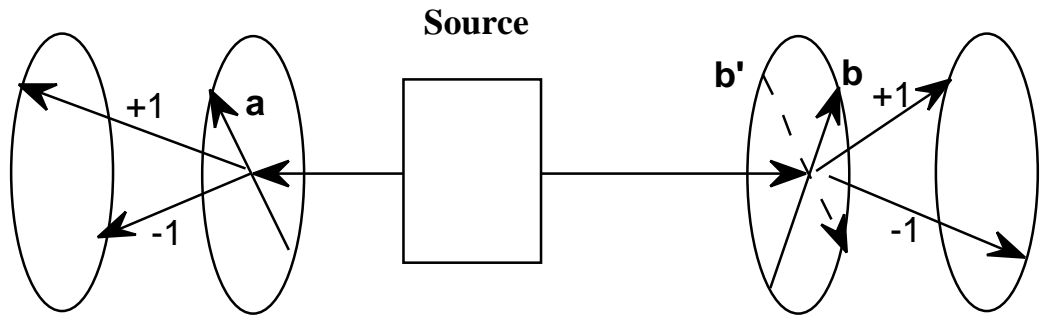

Fig. 1

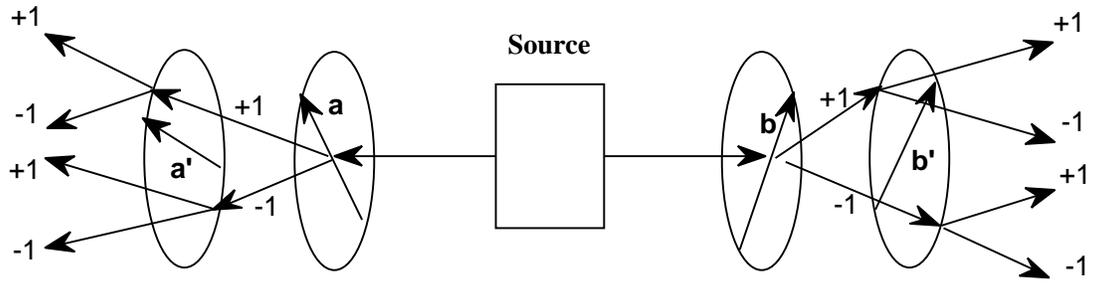

Fig. 2